\documentclass[acmlarge,nonacm]{acmart}

\AtBeginDocument{\settopmatter{printacmref=true}}
\makeatletter
\newcommand{\myconfshort}{\acmConference@shortname}
\newcommand{\myconffull}{\acmConference@name}
\newcommand{\myconfdate}{\acmConference@date}
\newcommand{\myconfloc}{\acmConference@venue}
\AtBeginDocument{
  \fancypagestyle{firstpagestyle}{
    \fancyhead{}%
    \fancyfoot[C]{}%
  }
  \fancyhf{}
  \fancyhead[LO]{\@headfootfont\shorttitle}%
  \fancyhead[RE]{\@headfootfont\@shortauthors}%
  \fancyhead[LE]{\@headfootfont\footnotesize \myconfshort, \myconfdate, \myconfloc}%
  \fancyhead[RO]{\@headfootfont\footnotesize \myconfshort, \myconfdate, \myconfloc}%
  \fancyfoot[C]{}%
}
\makeatother
\acmBooktitle{\conffull\@ (\confshort), \confdate, \confloc}

\usepackage{subcaption}
\AtBeginDocument{%
  }

\setcopyright{acmlicensed}
\copyrightyear{2026}
\acmYear{2026}
\setcopyright{cc}
\setcctype{by}
\acmConference[FAccT '26]{The 2026 ACM Conference on Fairness, Accountability, and Transparency}{June 25--28, 2026}{Montreal, QC, Canada}
\acmBooktitle{The 2026 ACM Conference on Fairness, Accountability, and Transparency (FAccT '26), June 25--28, 2026, Montreal, QC, Canada}
\acmDOI{10.1145/3805689.3806492}
\acmISBN{979-8-4007-2596-8/2026/06}

\usepackage{multirow}
\usepackage{array}

\usepackage{enumitem}
\begin{document}

\title{Mapping Emerging Climate Misinformation Playbooks in the Global South}


\author{Marcelo Sartori Locatelli}
\authornote{Both authors contributed equally to this research.}
\orcid{0000-0002-0893-1446}
\email{marcelo.sartori-locatelli@mpi-sp.org}
\affiliation{%
  \institution{MPI-SP}
  \city{Bochum}
  \country{Germany}
}
\affiliation{
\institution{UFMG}
\city{Belo Horizonte}
\country{Brazil}
}

\author{Wenchao Dong}
\authornotemark[1]
\orcid{0000-0001-5586-2335}
\email{wenchao.dong@mpi-sp.org}
\affiliation{%
  \institution{MPI-SP}
  \city{Bochum}
  \country{Germany}
}

\author{Pedro Loures Alzamora}
\email{pedro.loures@dcc.ufmg.br}
\affiliation{%
  \institution{UFMG}
  \city{Belo Horizonte}
  \country{Brazil}
}

\author{Pedro Dutenhefner}
\email{pedroroblesduten@ufmg.br}
\affiliation{%
  \institution{UFMG}
  \city{Belo Horizonte}
  \country{Brazil}
}

\author{Wagner Meira Jr.}
\authornote{Co-corresponding authors.}
\email{meira@dcc.ufmg.br}
\affiliation{%
  \institution{UFMG}
  \city{Belo Horizonte}
  \country{Brazil}
}

\author{Meeyoung Cha}
\authornotemark[2]
\orcid{0000-0003-4085-9648}
\email{mia.cha@mpi-sp.org}
\affiliation{%
  \institution{MPI-SP}
  \city{Bochum}
  \country{Germany}
}
\affiliation{
\institution{KAIST}
\city{Daejeon}
\country{South Korea}
}

\author{Virgilio Almeida}
\authornotemark[2]
\orcid{0000-0001-6452-0361}
\email{virgilio@dcc.ufmg.br}
\affiliation{%
  \institution{UFMG}
  \city{Belo Horizonte}
  \country{Brazil}
}

\renewcommand{\shortauthors}{Locatelli et al.}


\begin{abstract}
Climate misinformation continues to erode support for climate action, a challenge that is especially acute in the Global South, where high climate vulnerability intersects with development pressures. In rapidly evolving digital ecosystems, misinformation adapts to platform incentives, shifting from overt rejection of climate science toward more subtle narratives that contest proposed solutions.
This study integrates large‑scale platform data with qualitative content analysis to examine how information systems shape contemporary climate discourse. Using a dataset of 226,775 climate‑related YouTube videos from Brazil (2019–2025), we identify two dominant misinformation strategies: traditional denial that disputes scientific evidence and an emerging ``new denial'' that accepts climate change while undermining mitigation and adaptation policies.
We find a pronounced transition to solution‑focused narratives that target renewable energy, climate governance, and environmental advocates. New denial content is produced by a wider array of actors, attracts higher engagement, and employs more sophisticated persuasive techniques. These patterns disproportionately affect regions already facing structural inequities and bring broader concerns about platform accountability in unequal information environments and suggest the need for governance approaches capable of addressing new denial, a rapidly adapting form of harmful content that often evades existing moderation policies.

\end{abstract}




\maketitle

\section{Introduction}

Climate change is one of the most pressing challenges confronting humanity in the 21st century. 
As the climate crisis intensifies, it has become the site of an escalating information conflict in which multiple forms of denialism obscure scientific consensus and delay policy action.
Despite robust evidence that human activities are driving climate change~\cite{oreskes2004scientific}, contributing to increased extreme weather events~\cite{national2016attribution}, and growing threats to food security~\cite{duchenne2021climate}, denialist narratives continue to proliferate across social media~\cite{gounaridis2024social}.
These movements leverage digital platforms to simulate scientific legitimacy, amplifying coordinated disinformation campaigns that fragment public understanding~\cite{schroeder2025malicious}, erode trust in expertise~\cite{zhang2022shifting}, and recast climate change as a polarized political identity issue rather than a scientific one.

\begin{figure*}[tb]
    \centering
    \includegraphics[width=\linewidth]{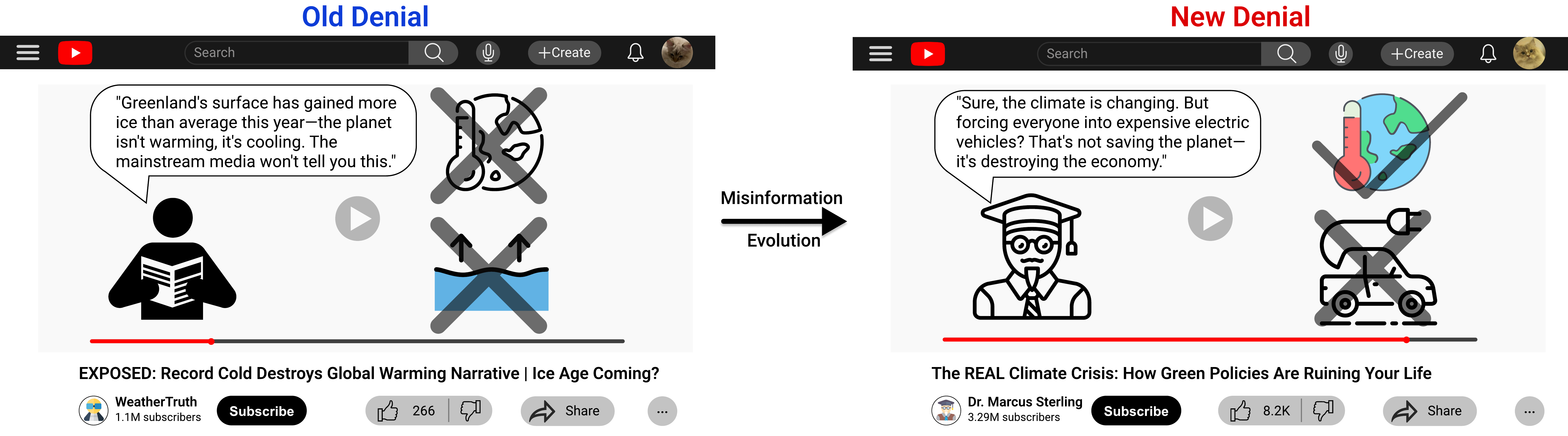}
    \caption{The evolution of denial narratives. \textit{Old Denial} rejects the existence of climate change, while \textit{New Denial} accepts climate change but attacks solutions, scientists, and the climate movement.}
    \label{fig:illustration}
    \Description[Illustrative examples of videos for old and new denial narratives.]{Illustrative examples of videos for old and new denial narratives. The old denial example has text related to the global cooling narrative, one of the topics seen on Brazilian YouTube. The new denial example is related to the narrative that climate solutions will ruin the economy.}
\end{figure*}

Climate change is also deeply entangled with longstanding global inequalities, and these disparities are reproduced and intensified within digital information ecosystems.
High-income countries disproportionately contribute to extreme events~\cite{schongart2025high}, while rising temperatures intensify poverty, particularly in underdeveloped regions~\cite{dang2025impacts}.
These same communities often encounter structural disadvantages in digital spaces: limited connectivity, uneven access to formal education, and historical exclusion from political institutions shape how people engage with online information and how resilient they can be to misleading or deceptive content~\cite{wilner2025rooted}. 

Within platformized environments, these inequalities become sociotechnical. Algorithms, moderation practices, and linguistic gaps shape who encounters climate misinformation and whose knowledge is marginalized. Disinformation campaigns exploit these vulnerabilities to deepen asymmetries in authority and participation, drawing on linguistic divides, low media literacy, and distrust rooted in colonial and postcolonial histories.
As a result, misinformation does not merely distort public understanding; it exacerbates environmental injustice by undermining adaptive capacity, skewing policy priorities, and reproducing global inequities in climate responsibility, risk exposure, and access to solutions.
Addressing these harms requires clarifying what constitutes credible evidence and interrogating the sociotechnical infrastructures that mediate climate knowledge, shape exposure to misinformation, and determine whose voices are legible within digital publics.

Combating climate misinformation cannot be a static approach as misinformation 
evolves in response to scientific and political progress.
Recent research has documented an important evolution in climate misinformation.
While traditional denial focused on outright denying the existence or anthropogenic causes of global warming~\cite{dunlap2010climate}, a ``new denial'' has emerged that accepts climate science, but attacks proposed solutions (e.g., renewable energy, climate activism, and environmental policies; see Figure \ref{fig:illustration} for an illustrative example)~\cite{ccdh2024denial, nicolosi2025new} . 
This shift is more insidious, as new denial can undermine climate belief even among those who accept the scientific consensus.
Without accurate, transparent, and diverse data sources, democratic decision-making and effective climate action are unattainable~\cite{klein2024data}.
Systematic observational data, including large‑scale data from the Web, are therefore essential for understanding how climate information circulates, who shapes public discourse, and how narratives influence behavior and policy outcomes.

In analyzing data, a balanced view is important, yet efforts to combat climate misinformation remain uneven, with most data concentrated on the United States and the broader Global North.
This geographic limitation matters for understanding global misinformation dynamics, as attitudes toward climate change vary widely across populations~\cite{vlasceanu2024addressing}.
Brazil represents a critical case: as host of COP30 in 2025, steward of the Amazon rainforest, and a major developing economy, the country occupies a pivotal position in global climate governance. 
Climate denial in Brazil has also been weaponized for political purposes, particularly to legitimize agricultural expansion and delegitimize left-leaning parties~\cite{miguel2022meada}. 
With YouTube reaching more than 67\% of Brazilians~\cite{DataReportal2025Brazil}, tracing how denial narratives evolve on this platform has significant implications for both local and global climate discourse.

We present the largest characterization of how climate denial evolved on Brazilian YouTube from 2019 to 2025. Drawing on a dataset of 226,775 videos, we advance the understanding of climate misinformation by:
\begin{itemize}[nosep]
    \item  \textbf{Longitudinal analysis of denial narrative evolution.} We document the temporal dynamics of old versus new denial in Brazilian YouTube, revealing that new denial has grown substantially since 2023. We identify key events and actors driving these shifts.
    
    \item \textbf{Multi-dimensional characterization of denial narratives.} We analyze how old and new denial differ across multiple dimensions: channel types producing each narrative, audience engagement patterns, persuasion strategies employed, and psychological responses elicited in comment sections. Our findings reveal that new denial is less personality-driven, generates higher engagement, and employs more sophisticated persuasion techniques combining emotional, logical, statistical, and moral appeals.

    \item \textbf{Implications for content moderation and platform accountability.} Our findings highlight a governance gap: new denial narratives that may be equally harmful to climate action often fall outside current platform policies, raising critical questions for platform accountability in the Global South context.
\end{itemize}

Our analysis reveals a concerning trajectory. 
The new denial in Brazil has expanded beyond individual content creators to independent digital media and traditional news outlets, reaching broader audiences through content that is both more engaging and more difficult to moderate. 
We argue that climate policy must prioritize evidence generation and systematic analysis, using empirical data to inform decisions while accelerating the production of new knowledge. 
Evidence-based climate policy, grounded in real-world data, including data from social media platforms, is essential to ensure that innovation advances climate solutions rather than undermining them.

\section{Background}

\subsection{Climate Science and Denial Narrative}

Climate change denialism and misinformation are concepts almost as old as the idea of climate change itself, emerging as a response to the increasing scientific consensus that human action has repercussions on global temperatures and the environment~\cite{dunlap2011organized}.\footnote{We use ``misinformation'' as an umbrella term to denote both misinformation (false information shared without intent to harm) and disinformation (false information deliberately disseminated to deceive)~\cite{wardle2017information}. Our research focuses on content analysis rather than inferring producer intent, thus, we do not further distinguish between these two types of false climate information dissemination.}
Early denial narratives claimed that no change in temperature was being recorded and, if it was, that it was natural. 
Over time, however, these narratives evolved and expanded to have a wide range of theories as to why the phenomenon is not real or cannot be prevented~\cite{hornsey2022toolkit}. 
These arguments were bolstered by the illusion of a lack of scientific consensus; in reality, a vast majority of scientists endorse anthropogenic global warming~\cite{cook2013quantifying}.

As denial strategies rapidly evolved, researchers sought to categorize the arguments made by those skeptical of human-induced global warming. 
Jeffrey Mazo~\cite{mazo2013climate}, inspired by the Skeptical Science blog, proposed four distinct categories, also called the four stages of climate change denial: it is not happening, it is not us, it is not bad, and it is too hard to fix. 
Cook~\cite{cook2022understanding} later refined this taxonomy, replacing the last category with two new ones: experts are unreliable, and climate solutions will not work. This resulted in five categories that parallel the five key climate beliefs~\cite{ding2011support}. 
This taxonomy formed the basis for the division between old and new denial introduced in the 2024 report by the Center for Countering Digital Hate (CCDH)~\cite{ccdh2024denial}. In this framework, old denial is defined as contradicting scientific consensus around climate change, represented by the first two categories of Cook's taxonomy: it is not happening and it is not us. 
In contrast, new denial centers on attacking, discrediting, and undermining climate solutions and their supporters. 
This shift is particularly concerning as new denial can be just as harmful as explicitly casting doubt on climate change, particularly given the narrative's substantial growth in popularity in English-speaking online communities over the past decade~\cite{nicolosi2025new}. 
Measuring the dynamics of these denial narratives is thus essential for understanding how misinformation ecosystems evolve and shape public discourse.

\subsection{Climate Discourse on Social Media }

Social media has become a critical arena for understanding the spread of climate misinformation.
The advent of social media networks amplified the impact of climate denial by fostering
polarization and facilitating the spread of misinformation across large numbers of users~\cite{treen2020online, falkenberg2022growing}.
Platform architecture plays a crucial role in this dynamic, where recommendation systems may induce and reinforce biases in people's information consumption~\cite{ye2025auditing}, while the structure of social media platforms helps the spread of denialist and conspiratory content~\cite{allgaier2019science}.
Misleading visualizations can also garner more engagement than accurate ones~\cite{lisnic2024yeah}.
Furthermore, research has shown that climate discussions often orbit around small groups of users, which use the content as campaign tools~\cite{shapiro2018climate}.

Analyzing climate discourse on social media is thus crucial to characterize how denial evolves and propagates.
Researchers have started to draw inspiration from Cook's taxonomy for analyzing social media data. Gounaridis and Newell~\cite{gounaridis2024social} characterize climate denial on Twitter (X) in the U.S., finding evidence of coordinated attacks to discredit climate science and activism, key characteristics of new denial. In a similar vein, CARDS~\cite{coan2021computer} is a framework for identifying denialist claims from documents fundamental for the computational analysis of YouTube conducted by CCDH~\cite{ccdh2024denial}, which uncovered that new denial made up 70\% of American denialist content in 2023, a drastic shift when compared to the late 2010s.
However, measuring denial narratives in the age of generative AI and large language models poses significant challenges, as AI-generated misinformation appears more credible than human-created content due to enhanced emotional expressions and detail~\cite{zhou2023synthetic}.

\subsection{Public Opinion Informed Policy-Making}

Understanding online discourse is critical as it directly informs climate policy design and communication strategies.
However, simply presenting correct climate science knowledge has proven to be ineffective to debunk misinformation and may backfire by reinforcing existing beliefs~\cite{konstantinou2025behavior}.
This can partially be explained by the fact that many people embrace climate skepticism as an
expression of ideological concern rather than a result of cognitive
reflection on evidence, necessitating indirect approaches~\cite{hornsey2022toolkit}.
The consequences for policy are substantial.
Existing studies of carbon taxes indicate that the public often mistakenly perceives climate policies as regressive and environmentally ineffective~\cite{douenne2022yellow}.
Public support for carbon pricing remains structured by partisanship and ideology, even when households receive material benefits~\cite{mildenberger2022limited}.
These findings suggest that traditional top-down knowledge dissemination could be insufficient, and taking a participatory approach that engages diverse stakeholders may lead to more public support~\cite{qi2025participatory}.

\subsection{Limited Attention for Global South}

Despite the global importance of analyzing climate change public opinions~\cite{vlasceanu2024addressing}, limited work has been done regarding this shift in the paradigms of climate skepticism specifically for lower resource communities, especially those in the Global South~\cite{tomassi2025disinformation}. 
This is concerning, as the characteristics of denial may change depending on the country and culture~\cite{nartova2022role}, and findings from an American context may not generalize to other contexts. 
Given that countries in the Global South, such as Brazil, play a critical role in global climate change mitigation~\cite{lovejoy2019amazon}, understanding how denial is expressed and evolves is essential to strengthen efforts in combating climate change globally.

Brazilian social media in particular has received relatively little attention in studies of social media perceptions of climate change compared to the U.S. and Europe~\cite{takahashi2025building}. Salles et al.~\cite{salles2023far} discuss the repercussions of one uniquely influent climate denial video, displaying how skepticism and conspiracies are instrumentalized by political groups. Dong et al.~\cite{dong2025characterizing} conduct a large scale analysis of engagement and the potential harms of AI for generating climate related content.
Our work contributes to addressing this research gap by providing the first systematic analysis of how climate denial narratives have evolved in Brazilian Portuguese YouTube content.

\begin{table}[t]
\centering
\caption{Overview of Brazilian Portuguese YouTube climate discourse dataset. Videos are annotated with persuasion strategies, while comments are labeled with Theory-of-Mind mental states. Channels represent the content creator sources.}
\label{tab:dataset_details}
\resizebox{.9\linewidth}{!}{%
\begin{tabular}{l|c|p{8cm}|p{4cm}}
\toprule
& \textbf{Size} & \textbf{Psychological label} & \textbf{Channel} \\
\midrule
\textbf{Video} & 226,775 & 
\textit{\textbf{Persuasion Strategy}} \newline
\quad \emph{Logical Appeal}: appealing with reasons and evidences \newline
\quad \emph{Emotional Appeal}: eliciting emotional feelings \newline
\quad \emph{Statistical Evidence}: providing concrete data, statistics \newline
\quad \emph{Social Norm}: creating pressure through social acceptance \newline
\quad \emph{Authority}: citing experts, institutions, and official reports \newline
\quad \emph{Personal Stories}: explaining individual experiences \newline
\quad \emph{Moral Appeal}: appealing with ethical responsibility \newline
\quad \emph{Reciprocity}: emphasizing mutual benefits of giving back \newline
\quad \emph{Scarcity}: presenting limited time and irreversible impacts \newline
\quad \emph{Common Ground}: building shared identity and values
& 
\multirow{2}{4cm}{Individual content creators \newline
Commercial companies \newline
Traditional news outlets \newline
Individual educators \newline
Independent digital media \newline
Non-profit organizations \newline
Formal education organizations \newline
Research institutes \newline
Local government \newline
National government \newline
Industry representatives \newline
Religious or spiritual org. \newline
International organizations \newline
Political party} \\
\cmidrule{1-3}
\textbf{Comment} & 2,756,165 & 
\textit{\textbf{Theory-of-Mind (ToM)}} \newline
\quad \emph{Belief}: information states that people hold to be true \newline
\quad \emph{Intention}: committed choices with planned actions \newline
\quad \emph{Desire}: motivational states representing preferences \newline
\quad \emph{Emotion}: affective states emotional responses \newline
\quad \emph{Knowledge}: organized representations of information \newline
\quad \emph{Percept}: sensory or socially shared perceptions \newline
\quad \emph{Non-literal}: using figurative or indirect language
& \\
\bottomrule
\end{tabular}
}
\end{table}

\section{Methodology}

\subsection{Dataset Characterization}

This research builds on the data set from previous research that analyzes psycholinguistic features of climate discourse~\cite{dong2025characterizing}.
The data set annotates and validates the persuasion strategies used in videos, capturing how climate messages are conveyed; for instance, using logical arguments to emphasize cause-and-effect relationships in climate science. 
To quantify the viewer responses, each comment was annotated with Theory of Mind (ToM) mental states such as beliefs about the legitimacy of climate change (\textit{belief}).
We obtained data from their publicly available repository~\footnote{\url{https://doi.org/10.5281/zenodo.17551955}}, which includes
226,775 videos and 2,756,165 comments
from Portuguese-language YouTube spanning from 2019 to 2025, and rehydrated it using the YouTube Data API V3 to get the video content.
Table~\ref{tab:dataset_details} details the persuasion strategies for videos, ToM mental states for comments, and channel categorizations.\footnote{Following established research practices in YouTube misinformation studies~\cite{ribeiro2020auditing}, we analyze aggregated channel-level statistics to identify systematic patterns in denial content production rather than targeting individual creators.}

To capture the semantic meanings of climate videos, we combined their titles, descriptions, and transcripts (when available) into unified textual representations. 
We generated embeddings using OpenAI's \texttt{text-embedding-3-small} model~\cite{openai2024embedding}, which transforms video content into numerical representations in 1,536 dimensions. 
Since the embedding model supports a maximum context length of 8,192 tokens, we truncated the combined textual descriptions for longer videos (7\% of the dataset) to fit within this limit.

\subsection{Climate Stance Annotation}

\begin{table}[bt]
\centering
\caption{Climate narrative definitions and examples from YouTube videos (English translations in parentheses).}
\resizebox{\linewidth}{!}{%
\label{tab:climate_narrative_definition}
\begin{tabular}{p{2cm}p{10cm}p{5cm}}
\toprule
\textbf{Narrative} & \textbf{Definition} & \textbf{Example Video Title} \\
\midrule
\textbf{New Denial} & Acknowledgment of climate change existence while minimizing its severity, questioning the reliability of climate science, dismissing the feasibility of proposed solutions, or arguing that impacts are beneficial. & \textit{Querem fazer um lockdown climático} (They want to implement a climate lockdown) \\
\midrule
\textbf{Old Denial} & Outright rejection of anthropogenic climate change, including claims that global warming is not occurring, that greenhouse gases do not cause warming, or that observed changes represent natural variation. & \textit{A Grande Farsa do Aquecimento Global} (The Great Global Warming Hoax) \\
\midrule
\textbf{Other} & Climate-related content that acknowledges or does not reject the consensus on climate change without employing denialist rhetoric, such as policy discussions, technical mitigation strategies, and scientific findings. & \textit{Por que está fazendo tanto CALOR no Brasil e no mundo?} (Why is it so HOT in Brazil and around the world?)  \\
\bottomrule
\end{tabular}
}
\end{table}

We ground our analyses in the CCDH report~\cite{ccdh2024denial} and define old denial as content related to two categories of Cook's taxonomy, it is not happening and it is not us, and new denial as content relating to the remaining three: it is not bad, solutions will not work, and experts are unreliable. 
Table~\ref{tab:climate_narrative_definition} presents the definitions and video examples of these two narratives, along with the definition of other non-denial climate content. 
To understand how climate denialism evolves on Brazilian YouTube at scale, we employ text classification techniques to determine each video's stance on climate change and its associated narratives. 
We develop an approach based on Chain-of-Thought (CoT) prompting with GPT-4.1-mini. For each video's title, description, and transcript, we assess whether the content contains ``old denial,'' ``new denial,'' or ``other'' narratives regarding climate change. Since our primary interest is denialist content, ``other'' serves as a catch-all label, mutually exclusive to the denial categories, including videos that express belief in anthropogenic climate change or remain neutral.

To evaluate classifier performance, two authors independently annotated 351 randomly sampled videos. 
Given the initial distribution's heavy skew toward ``other'', we created an additional balanced set of 351 videos using Qwen3-30B to generate pseudo denial labels, ensuring adequate representation of denial narratives in our evaluation set for a final total of 702 human-annotated videos, covering 13 of 14 channel categories across all seven years.
We measure inter-annotator agreement using Cohen's kappa, obtaining a score of 0.89 for denial detection and 0.87 for specific narrative classification. Disagreements were solved with open discussion for creating the final evaluation set.
The classifier achieves strong performance in detecting denialist narratives overall (accuracy = 0.90, F1 = 0.86, weighted F1 = 0.90), though distinguishing between old and new denial proves more challenging (accuracy = 0.77, F1 = 0.75, weighted F1 = 0.77). 
This performance difference reflects the nuanced distinctions between these categories, which can be subtle even for human annotators. 
Importantly, our approach is effective and scalable for annotating the entire YouTube dataset.
All annotation details are provided in Appendix~\ref{appendix:prompt}.

To investigate the characteristics of denialist content in Brazil and contrast the more established old denial with the emerging new denial, we identify key topics of discussion in both categories of content through the use of the BERTopic~\cite{grootendorst2022bertopic} algorithm in order to unveil what the main themes of discussion are. Each topic was manually labeled from the keywords identified by the algorithm and a video sample. Additionally, we identify the entities cited in each video using the \texttt{Spacy} library~\cite{Honnibal_spaCy_Industrial-strength_Natural_2020}, which are useful in determining exactly who and what are central to denialist content on YouTube.

\begin{figure*}[t]
    \centering
    \includegraphics[width=\linewidth]{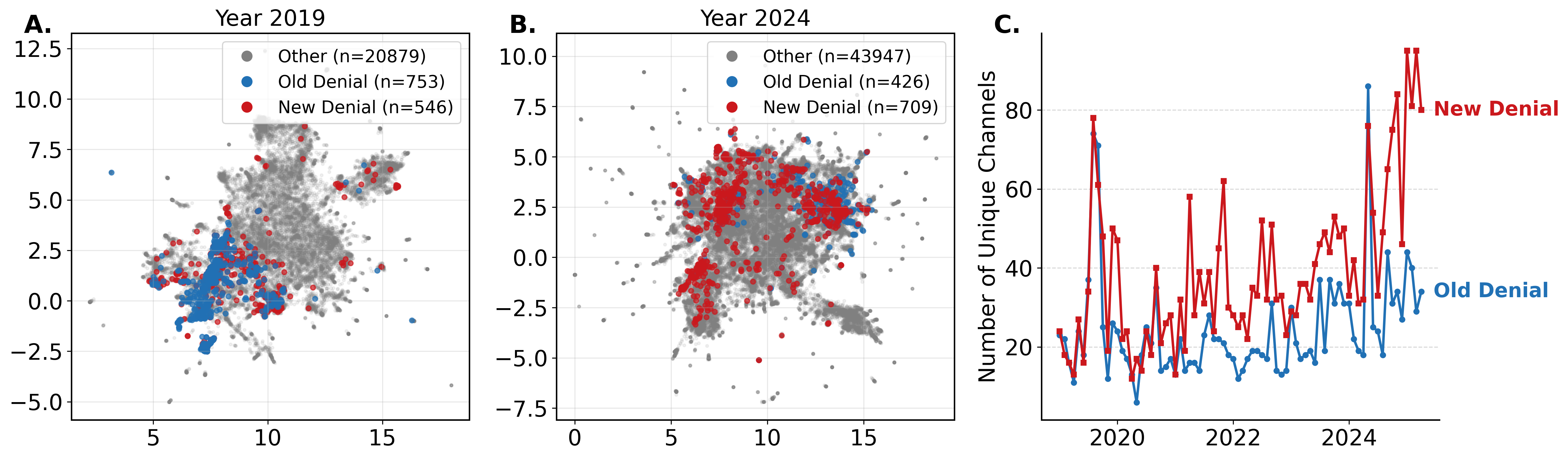}
    \caption{UMAP visualization of video content embeddings for 2019 and 2024 (A-B), showing the semantic clustering of old denial (blue), new denial (red), and other climate-related content (gray). Each point represents a video, with outliers (z-score > 3) removed for clarity. (C) Monthly time series of unique YouTube channels producing old denial versus new denial content from 2019 to 2025.}
    \label{fig:denial_analysis_stats}
    \Description[New denial became more different from old denial over time.]{Video embeddings for 2019 show that old and new denial were initially similar, but by 2024 new denial had changed remarkably.}
\end{figure*}

\section{Results}

\subsection{Growing New Denial Narrative}

We observe a significant expansion of new denial content on YouTube, both in terms of volume and semantic diversity.

Fig.~\ref{fig:denial_analysis_stats}A and B illustrate semantic distributions of climate videos for 2019 and 2024, respectively.
In 2019, old denial (blue, n=753) and new denial content (red, n=546) exhibit substantial overlap within the embedding space, indicating that both denial strategies were discussing semantically similar themes.
This co-localization suggests that new denial narratives had not yet differentiated themselves substantially from traditional denial narratives.
By 2024, however, a pronounced shift emerges.
Although the old denial has decreased in volume (n=426), the new denial videos have expanded considerably (n=709).
Furthermore, the spatial distribution of the new denial content in 2024 suggests a greater semantic dispersion across the embedding space compared to 2019.
The progression across years reveals the shifting landscape of denial discourse, with old denial maintaining relatively stable semantic patterns, while new denial emerges and evolves, occupying diverse semantic spaces that become increasingly prominent in later years.
For complete yearly embedding distributions from 2019 to 2025, see Figure~\ref{fig:app_denial_all_embeddings} in the Appendix.

Fig.~\ref{fig:denial_analysis_stats}C provides complementary evidence through a monthly time series summarizing the number of unique YouTube channels that produce each type of denial content from 2019 to 2025.
The trajectories show relatively comparable levels of channel activity between old and new denial during the early years.
However, a considerable shift emerges around 2021, after which channels producing new denial content outpace those producing old denial.
Mann-Kendall trend tests confirm statistically significant increasing trends for both denial types, though the trajectories differ substantially in magnitude.
New denial channels exhibit a moderate-to-strong upward trend ($\tau = 0.45$, $p < 0.001$) with a Sen's slope of 0.49 channels per month, while old denial channels show a weaker increase ($\tau = 0.23$, $p < 0.01$) with a Sen's slope of 0.14.
This indicates that new denial channels are growing at a rate 3.5 times higher than old denial channels.
This divergence implies that an increasing number of YouTube content creators have become involved in disseminating climate misinformation through new denial strategies.
This also suggests a broadening of the base that produces misinformation rather than simply increasing the output of existing channels.

\begin{figure*}[tb]
    \centering
    \includegraphics[width=\linewidth]{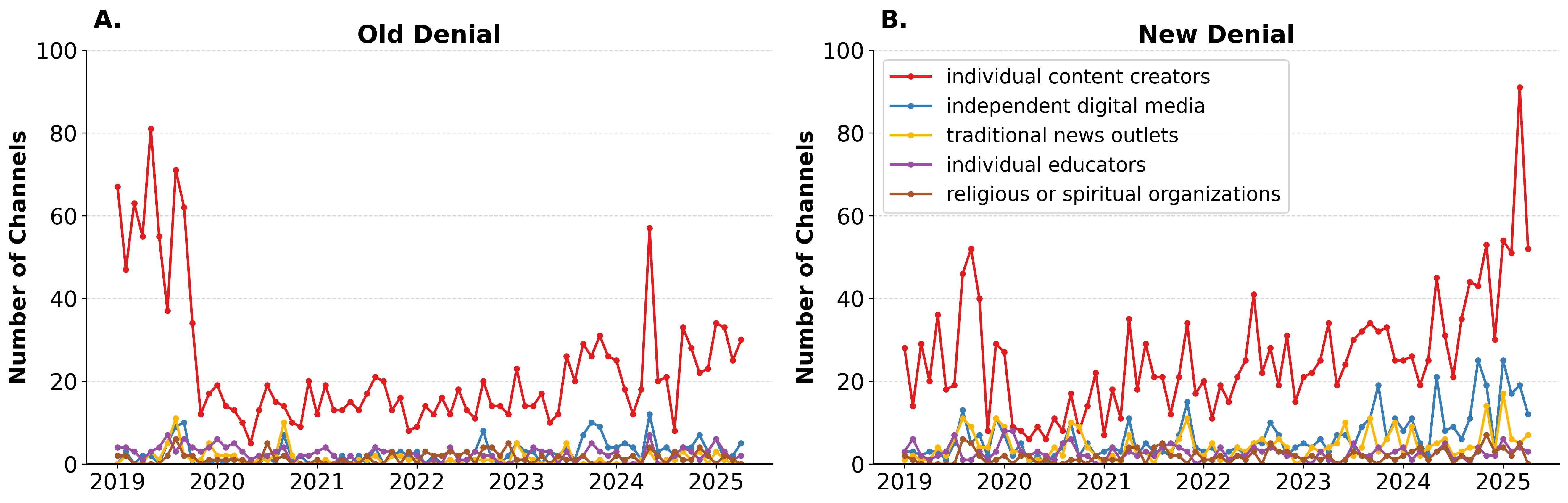}
    \caption{Monthly distribution of YouTube channels producing (A) old denial content and (B) new denial content, by channel type from 2019 to 2025. The top five channel categories by volume are included for clarity.}
    \label{fig:denial_channel_distribution}
    \Description[Distribution of channel categories posting denialist content. Individual content creators make up the majority of channels.]{Distribution of channel categories posting denialist content. Individual content creators make up the majority of channels. The number of channels posting new denial is growing, while old denial is decreasing over time.}
\end{figure*}

Across both types of denial, individual content creators dominate climate misinformation production on YouTube.
Fig.~\ref{fig:denial_channel_distribution}A shows that the old denial activity of individual content creators peaked in 2019 with more than 80 channels per month, followed by a sharp decline after early 2020. 
It stabilized at lower levels with fewer than 20 channels publishing each month until 2023, from which point it shows a gradual increase. 
All other four types of channels remained stable with fewer than 10 channels per month. 
In contrast, as shown in Fig.~\ref{fig:denial_channel_distribution}B, individuals producing new denial content demonstrate sustained growth over time, intensifying from late 2023 and reaching approximately 90 channels per month by early 2025. 
Furthermore, independent digital media and traditional news outlets also exhibit notable increases in new denial content from 2023. 
This pattern suggests that the expansion of new denial narratives extends beyond grassroots creators to more institutionalized media actors, confirming the broadening of the information-producing base.

\subsection{Themes of Brazilian Climate Denial}

\begin{figure*}[tb]
    \centering
    \includegraphics[width=\linewidth]{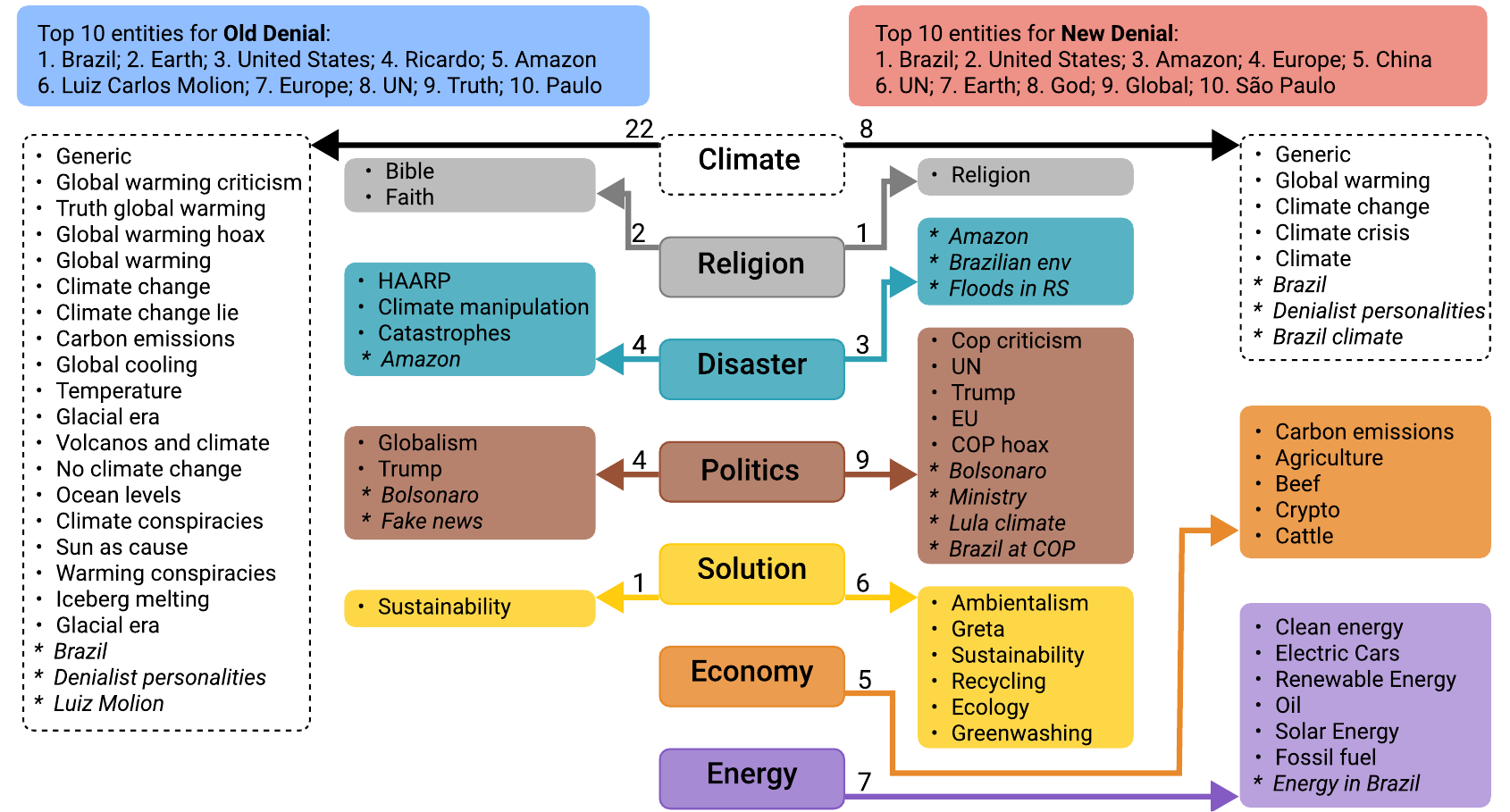}
    \caption{Thematic structure of old (left) and new (right) denial topics, organized into seven major themes. Numbers indicate topic counts per theme. Bullet points (\textbullet) denote global topics appearing in international climate denial discourse; asterisks (*) indicate Brazil-specific topics. Top panels show top 10 named entities for each narrative.}
    \label{fig:topics}
    \Description[Identified topics for old and new denial.]{Identified topics for old and new denial. Each topic is categorized as relating to either religion, politics, disasters, solutions, economy, energy or climate. New denial topics are more well distributed among categories.}
\end{figure*}

Fig.~\ref{fig:topics} shows an overview of the identified entities and topics by denial category, which we group into high level themes. 
New denial spans 39 topics across seven themes, including economy and energy, which are absent in old denial.
From this, we can see that there is a large discrepancy in the variety of topics covered by the narratives as, while old denial is mostly restricted to discussions explicitly about climate or temperatures, new denial is more pervasive, covering a large array of issues, spanning proposed solutions for climate change, energy generation, farming, among others. 
The differences become even clearer when we compare the top entities cited by the videos in each category. Old denial content is often associated with notorious Brazilian climate denialists: Ricardo (Felício), and Luiz Carlos Molion~\cite{miguel2022meada}, or known conspiracies, such as the new world order. Meanwhile, new denial displays its less centralized nature, with the only mentions of people among the top most cited entities being Brazilian politicians Lula and Bolsonaro, who either are president or were president at some point during the studied period, justifying their high presence. This may be one of the reasons for the lower variety of old denial topics, as many of the videos in our dataset are driven by speeches and interviews of key personalities (Ricardo Felício, Luiz Molion, Olavo de Carvalho, Dom Bertrand, to name a few among the top entities), leading to a more constant stream of videos that may not be as affected by specific events, as we will see later in this section.

\begin{figure*}[tb]
    \centering
    \includegraphics[width=\linewidth]{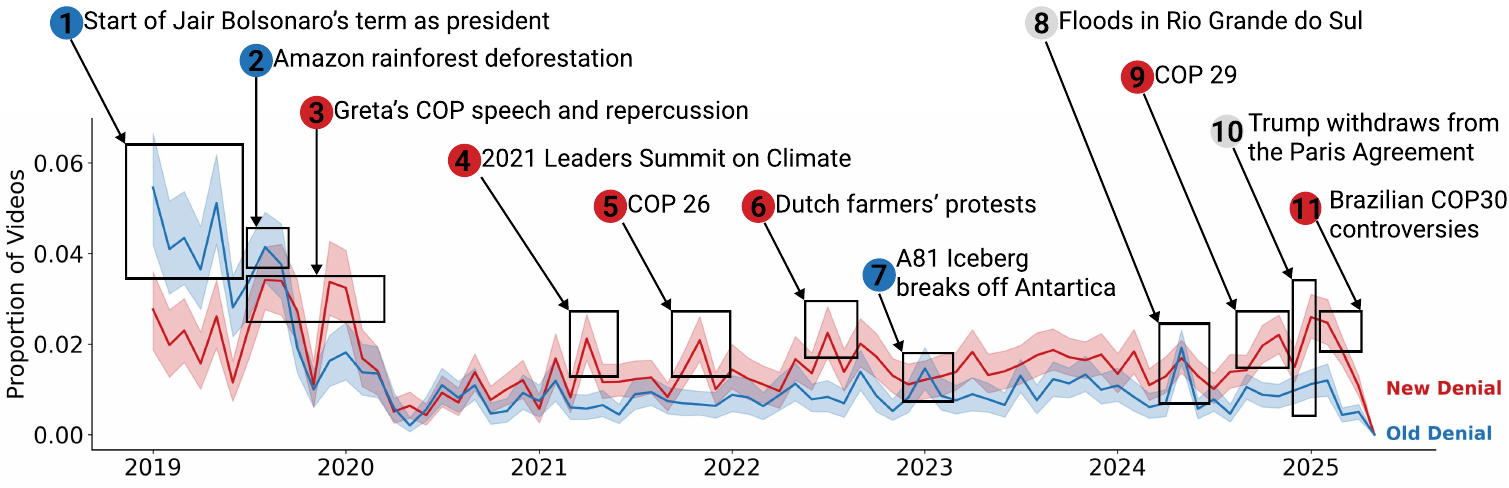}
    \caption{Proportion of videos related to old and new denialist narratives over time. Events that received significant coverage by either groups are highlighted. The shaded area shows the 95\% bootstrapped confidence interval for the proportion.}
    \label{fig:timeline}
    \Description[Proportion of new and old denial videos per month. Old denial is more prevalent before 2020 and new denial after.]{Proportion of new and old denial videos per month. Old denial is more prevalent before 2020 and new denial after. There are 11 highlighted relevant political or climate related events that are correlated with increased number of denial videos.}

\end{figure*}

Fig.~\ref{fig:timeline} shows the proportion of new and old denial videos over time. 
Using dynamic topic modeling through BERTopic, we track how topic distributions shift over time and identify events associated with sharp increases in each category (events 1 to 11).
 
These labels do not imply linear replacement; old denial persists as a background narrative throughout, and the two forms consistently overlap.
Our dataset starts from the first year of Bolsonaro's presidency, which is reflected in our data by a big peak in old denialist narratives from the beginning (event 1 and 2). 
In this case, Bolsonaro's denial of Brazil's role in the destruction of the Amazon and its repercussions for the environment~\cite{ferrante2019brazil}, together with the many statements by his supporters fueled a large number of video publications on the early stages of 2019, such as \textit{``Senator Bittar points out the hypocrisy and manipulation of those who attack Bolsonaro regarding the Amazon and global warming.''} containing the statement \textit{``Everything is done in the name of combating anthropogenic global warming, caused by humans, which has already been widely refuted''}.

The same level activity was not initially seen for new denial, which targets proposed solutions for climate change related problems. However, when Greta was brought to the forefront after her UN speech (event 3), many saw an opportunity to discredit climate activism~\cite{dave2022targeting}, as attributes such as her perceived lack of credentials, and unclear backing were frequently criticized by users online, sparking a plethora of Brazilian videos wishing to capitalize on these narratives, such as \textit{``Greta really is a farse''} or \textit{``Who is behind Greta Thunberg?''}, comprising from 20 to 40\% of new denial videos on the highlighted peaks. The contrast between this and the previous two events perfectly exemplifies the differences between old and new denial as, unlike in topic Bolsonaro, where the existence of negative impacts related to anthropogenic global warming was being attacked, here, the focus is changed to the people trying to solve climate change, hence, we see quotes such as \textit{``Globalists take children and use them as spearheads for movements ... this girl does not know what she's talking about''}. 
Similar discrediting of solutions followed during COP26 (event 5).

Another expression of new denial would be seen in 2022 with the farmers' protests in the Netherlands (event 6). There, instead of being discredited, the protestants, together with the increase in coal usage throughout Europe in that year, were seen as proof of failure of measures to reduce emissions, serving as a convenient proxy through which solutions could be criticized, a key characteristic of new denial narratives: \textit{``Outrageous! Farms will be closed to reduce gas emissions in the Netherlands''}.

Some events, however, are relevant to both groups, leading to a large amount of published videos that transcend the boundaries of old and new denial. This was the case with the floods in Rio Grande do Sul, Brazil, in 2024, whose destructive power was so great that it is considered one of the worst climate disasters in the country~\cite{marengo2024maior}. In this case, both of the denial narratives are associated with attempts to minimize the impact that climate change may have had on rains leading to the disaster, but they differ drastically in their approach. In old denial, videos such as \textit{``Climate change is natural and happens independently of human actions ok?''} published in response to the event attribute the abnormal volume of rain entirely to natural changes, while in  new denial, videos such as \textit{``Is Global Warming to Blame? Floods in Rio Grande do Sul''}, acknowledge anthropogenic climate change but attribute the disaster solely to other factors. During this period, conspiracies were also prevalent, namely the climate control conspiracy, which position the floods as an intentional disaster created by the High Frequency Active Auroral Research Program (HAARP), further minimizing any possibility that global warming may have played a role and, thus, being attributed to old denial.

This difference in treatment for the same topic was also evident after President Trump's inauguration and withdrawal from the Paris Agreement in 2025. On the old denial side, his position and statements on climate change inspired a relatively small number of videos. However, the influence of such decision is much greater on the new denial narratives as the decision to withdraw was used to support arguments discrediting the Paris Agreement and ESG (Environmental, Social, and Governance) measures.

\subsection{Climate Denial Persuasion and Reaction}

\begin{figure*}[t]
    \centering
    \includegraphics[width=\linewidth]{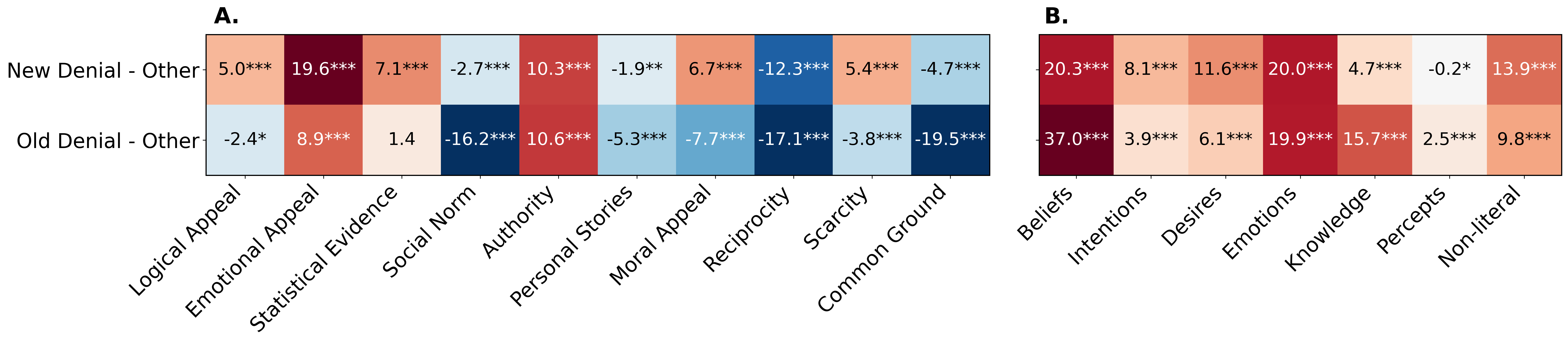}
    \caption{Difference in prevalence (\%) of (A) persuasions in video content and (B) ToM mental states in comments between denial and other climate-related content. Values represent percentage point differences compared to other climate content, with positive values (red) indicating higher prevalence in denial content and negative values (blue) indicating lower prevalence. Statistical significance based on chi-square tests with Benjamini-Hochberg correction: * $p < 0.05$, ** $p < 0.01$, *** $p < 0.001$.}
    \label{fig:denial_persuasion_tom}
    \Description[Heatmap with prevalence of persuasion strategies in video content.]{Heatmap with prevalence of persuasion strategies in video content. Authority and emotional appeal are more prevalent in both denial narratives when compared to other content. New denial employs more logical and moral appeals, statistical evidence and scarcity. }
\end{figure*}

Denial narratives employ distinct persuasion strategies and elicit different levels of psychological reactions.
Fig.~\ref{fig:denial_persuasion_tom}A
shows the different use of persuasions between denial and other climate content.
Both old and new denial rely significantly more on emotional appeal (new: $+19.6\%$, $\chi^2 = 604.30$; old: $+8.9\%$, $\chi^2 = 88.02$; both $p<0.001$) and authority (new: $+10.3\%$, $\chi^2 = 145.48$; old: $+10.6\%$, $\chi^2 = 111.26$; both $p<0.001$) compared to other climate content.
Furthermore, 
new denial employs significantly more 
logical appeal ($+5.0\%$, $\chi^2 = 33.63$, $p<0.001$),
statistical evidence ($+7.1\%$, $\chi^2 = 98.92$, $p<0.001$), scarcity ($+5.4\%$, $\chi^2 = 64.03$, $p<0.001$), and moral appeal ($+6.7\%$, $\chi^2 = 78.04$, $p<0.001$), whereas old denial use similar or lower levels of these persuasions compared to other general climate content.
These patterns suggest that new denial adopts a more sophisticated strategy with significant emphasis on logic, emotion, and morality.

Fig.~\ref{fig:denial_persuasion_tom}B examines mental state expressions in comments on denial videos.
Both denial types elicit significantly more psychological engagements from audiences across nearly all ToMs.
In particular, emotional expressions are significantly increased (new: $+20.0\%$, $\chi^2 = 32773.82$; old: $+19.9\%$, $\chi^2 = 14689.33$; both $p<0.001$).
While old denial videos prompt audiences to react with more knowledge-related responses ($+15.7\%$, $\chi^2 = 14065.85 $, $p<0.001$), new denial tends to elicit more non-literal expressions such as sarcasm or irony ($+13.9\%$, $\chi^2 = 17547.02$, $p<0.001$).
Finally, belief-related comments are significantly higher in denial videos (new: $+20.3\%$, $\chi^2 = 63847.78$; old: $+37.0\%$, $\chi^2 = 97114.47$; both $p<0.001$); however, the effect for new denial is moderate, suggesting an elevated yet more subtle influence pattern.

\subsection{Climate Denial Engagement}

\begin{figure*}[tb]
    \centering
    \includegraphics[width=\linewidth]{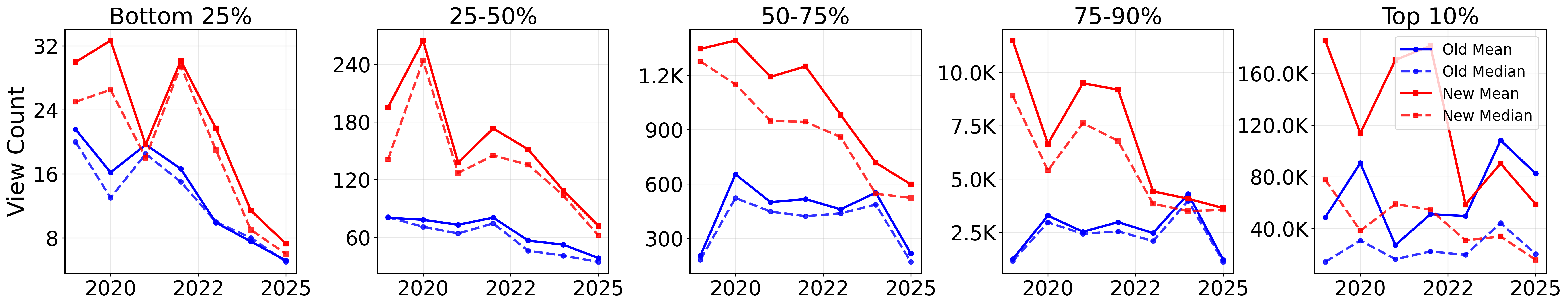}
    \caption{Yearly engagement measured by view count for old denial (blue) and new denial (red) videos across quantile groups. Solid lines represent mean values and dashed lines represent median values. See Fig.~\ref{fig:denial_engagement_like_comment} for like and comment metrics with consistent patterns.}
    \label{fig:denial_engagement}
    \Description[Engagement metrics for denial.]{Engagement metrics for denial. New denial metrics are generally higher than old denial.}
\end{figure*}

New denial content attracts more engagement from YouTube audiences compared to old denial information.
Fig.~\ref{fig:denial_engagement} presents engagement metrics across 
popularity groups (based on count quantiles)
for both denial types.
New denial videos consistently outperform old denial across all metrics and popularity groups, with particularly pronounced differences among the top 10\% most viewed groups.
The most popular new denial videos reach more than 150,000 viewers compared to approximately 50,000--100,000 for old denial.
The gap between mean and median values indicate a skewed distribution driven by a subset of highly viral content.
Note that both denial types show declining engagement trends toward 2025; however, this pattern reflects the recency of publication rather than diminishing audience interest, as newly published videos have had less time to accumulate views, likes, and comments. 
These findings suggest that new denial narratives not only reach broader audiences, but also generate substantially greater user interaction than traditional denial content.

\section{Discussion}

\subsection{Persuading Climate Misinformation}

New denial employs increasingly data-driven strategies to disseminate misinformation. 
Although conventional wisdom for combating misinformation emphasizes providing accurate information to counter false claims~\cite{nyhan2021backfire}, our results reveal a critical paradox: new denial videos present significantly more statistics and employ more logical appeal, including scarcity-based persuasion strategies to emphasize urgency (Fig.~\ref{fig:denial_persuasion_tom}). 
When compared to old denial narratives, which rely on less detailed content, new emerging denial videos garner substantially higher engagement (Fig.~\ref{fig:denial_engagement}). 
These findings align with Konstantinou and Karapanos's observation that misinformation interventions frequently fail to achieve their objectives despite employing detailed statistics~\cite{konstantinou2025behavior}. 
When climate denial narratives adopt the same presentation strategies that characterize legitimate climate science communication (e.g., detailed statistics and data visualizations), the general public faces increased difficulty distinguishing credible information from sophisticated misinformation.

Indeed, presenting scientific data alone is insufficient and can create vulnerabilities.
New denial narratives strategically select cherry-picked statistics and framing to present misleading content, which can be further amplified, as misleading data visualizations often garner greater engagement than accurate ones~\cite{lisnic2024yeah}.
Climate skeptics typically arrive at their conclusions based on ideology rather than evidence evaluation, then selectively interpret and present evidence to rationalize their pre-existing beliefs~\cite{hornsey2022toolkit}.
Our research echoes previous work suggesting that traditional misinformation exerts limited influence on social media compared to skeptical information: flagged misinformation undergoes fact-checking and debunking, resulting in lower exposure, while skepticism narratives evade verification and exert substantially greater overall influence~\cite{allen2024quantifying}.
These findings suggest the need for narrative detection and modeling approaches that extend beyond traditional stance detection or fact-checking methods.

In addition to identifying the persuasion strategies employed by the denial videos, we also identify the reactions of the critical audience through mental states (Fig.~\ref{fig:denial_persuasion_tom}).
Measuring psychological influences of climate content reveals that people may hold skepticism toward climate solution effectiveness, which is not denying climate change's existence but expressing pessimistic beliefs about policy effects~\cite{douenne2022yellow}.
Furthermore, developing effective misinformation interventions requires psychological understanding of how people can be motivated and influenced~\cite{sinclair2025behavioral}.
By examining both the rhetorical strategies of denial narratives and their psychological impacts on audiences, our work provides a foundation for designing interventions that address not only factual accuracy but also the persuasive mechanisms through which climate denial misinformation shapes beliefs.

\subsection{AI and Climate Misinformation}

Online climate discourse has been greatly influenced by AI systems\footnote{We here primarily discuss AI's influence on online climate discourse, but AI's impacts extend beyond digital spaces. AI development is increasingly affecting the physical environment, with the Global South experiencing more adverse impacts. See discussions, for example,~\cite{klein2024data, luccioni2025efficiency, valdivia2025data}.}, especially recent generative AI such as large language models (LLMs). With the advent of generative AI models, it is now easier than ever to create convincing human-like text~\cite{dong2025characterizing}, which opens new opportunities for combating climate misinformation at scale. Language models can be deployed to debunk climate misinformation~\cite{czarnek2025addressing}, while automated detection systems can identify and flag climate misinformation on social media~\cite{rojas2024hierarchical}. Furthermore, personalized AI systems can be tailored to individual users' needs to identify misinformation~\cite{luccioni2025efficiency}, and hybrid approaches enable humans and AI to collaboratively identify misinformation by combining algorithmic efficiency with human oversight~\cite{schmitt2024role}. These advances offer promising possibilities to address climate misinformation problem that has long undermined public understanding.
Our methodology uses LLM-based classification to characterize denial narratives across a large corpus, showing how AI can support systematic analysis of misinformation dynamics.

\begin{figure}[t]
    \centering
    \includegraphics[width=\linewidth]{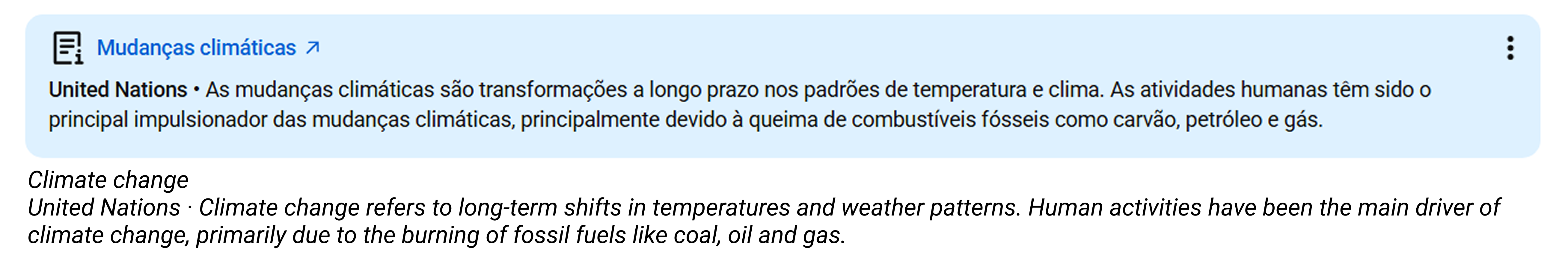}
    \caption{Information panel for Brazilian YouTube climate change videos. The English translation is shown below.}
    \label{fig:info_panel}
    \Description[Information panel for Brazilian YouTube climate change videos.]{Information panel for Brazilian YouTube climate change videos. In English: ``Climate Change United Nations · Climate change refers to long-term shifts in temperatures and weather patterns. Human activities have been the main driver of climate change, primarily due to the burning of fossil fuels like coal, oil and gas.''}
\end{figure}

At the same time, AI introduces new risks that contextualize and compound the trends documented in our analysis. 
AI makes misinformation generation easier and costless, and AI-generated misinformation is ranked as the second-highest global risk (53\%) by the World Economic Forum's 2024 Global Risk Report, just behind extreme weather (66\%)~\cite{wef2024globalrisks}. 
Research has shown how massive tailored climate misinformation can be easily automatically generated through language models~\cite{dong2025characterizing}.
What makes AI-powered misinformation propagation difficult to curb is that people often struggle to distinguish human-written text from AI-generated content~\cite{jakesch2023human}. AI-generated misinformation exhibits more emotional expressions and enhanced details than human-created content, making it appear more credible~\cite{zhou2023synthetic}. 
In addition to the challenges of AI-generated misinformation,
AI can compromise the reliability of misinformation detection.
Content moderation systems can make inconsistent decisions depending on algorithmic features~\cite{gomez2024algorithmic}, while AI-based fact-checking systems risk disproportionately benefiting majority communities~\cite{neumann2023does}, raising concerns about distributive injustice~\cite{neumann2022justice}.
Navigating such tensions while reaping the benefits of these emerging technologies will be an important challenge for the foreseeable future.

\subsection{Platform Accountability for Misinformation Moderation}

The spread of ``new denial'' demonstrates how engagement-driven platform incentives shape information visibility; this has been identified as a source of sociotechnical harm when algorithms systematically expose users to low-quality information and reinforce social biases~\cite{ribeiro2020auditing, ye2025auditing, shelby2023sociotechnical}.
YouTube has implemented several measures to address misinformation.
In 2019, the platform introduced fact-checking information panels in India and Brazil, later expanding them to the United States in 2020~\cite{YouTube2020FactChecks}.
These panels automatically appear on videos addressing sensitive topics, including climate change, providing users with information from authoritative sources (see Fig.~\ref{fig:info_panel}).
Research suggests that such credibility indicators can effectively reduce the sharing of false content~\cite{yaqub2020effects}, and recent work has explored more sophisticated systems for introducing credibility signals to video-sharing platforms~\cite{hughes2024viblio}. 
However, the coverage of these panels remains insufficient~\cite{godinez2024youtube}.

Our findings suggest that new denial content falls outside the scope of current moderation.
YouTube's climate misinformation policy explicitly targets only content that ``contradicts well-established scientific consensus around the existence and causes of climate change''~\cite{google2021climate}, while debate or discussion of climate change topics, including public policy or research, is allowed. 
As the CCDH report notes, YouTube confirmed that the majority of new denial videos ``did not breach their policies''~\cite{ccdh2024denial}. 
This creates a possible distributive injustice~\cite{neumann2022justice}, where users seeking climate information are systematically exposed to misleading new denial content.
Translating research insights into actionable platform features requires using computing's ability to measure social problems and diagnose how they manifest in technical systems~\cite{abebe2020roles}. 
Such diagnostics are essential for ensuring platform rules keep pace with evolving misinformation tactics, preventing harmful content from evading detection by design and distorting public understanding of a central global threat.

\subsection{Brazil Context and Policy Implications}

Brazil serves as our representative case from the Global South given its prominent role in global climate politics, its geographic significance to the Amazon rainforest, and the substantial volume of available climate-related online data. We recognize that the Global South includes diverse nations with different climate contexts, and this study does not intend to diminish that diversity.
The governance gap identified in our analysis intersects with a fundamental tension in Brazilian law: balancing free expression (Article~5) with the right to an ecologically balanced environment (Article~225).
Over-moderation risks participatory injustice by silencing legitimate debate, while under-moderation creates distributive injustice where users seeking climate information are systematically exposed to misleading new denial content~\cite{neumann2022justice}. Rather than content removal, we advocate for platform transparency and evidence-driven governance that can navigate this balance.

Brazil has pursued platform accountability through legislative proposals such as PL~2630/2020, which aims to enhance platform responsibility for misinformation~\cite{Brasil2020PL2630}. 
These efforts were further emphasized by the recent \textit{Statement of Commitment on Climate Information Integrity} launched at COP30, highlighting the joint effort between government and civil society to strengthen a more transparent, reliable, and resilient information ecosystem capable of addressing climate change~\cite{COP2025Brazil}. 
These regulatory efforts coincide with public consensus: 81\% of Brazilians believe social media platforms should be responsible for preventing the spread of fake news~\cite{datasenado2024misinfo}, and research confirms that users predominantly hold platforms accountable for failing to prevent misinformation~\cite{lima2022others}. 
Internationally, the European Union's Digital Services Act (DSA) represents a comprehensive regulatory framework that requires Very Large Online Platforms to identify and mitigate systemic risks, including the algorithmic amplification of misinformation~\cite{EuropeanCommission2025DSA}. 
The structural nature of algorithmic amplification requires such regulatory intervention beyond platform self-governance, as algorithmic systems are the product of a series of stakeholders~\cite{metcalf2023taking} whose broader influence on the general public demands accountability.
However, translating these commitments into practice remains challenging because existing frameworks primarily address overt misinformation and may be ill-equipped to handle emerging forms of ``new denial.'' These more subtle narratives necessitate nuanced regulatory approaches that enforce strict platform accountability to counter the evolving tactics of misinformation campaigns.

\subsection{Limitations and Future Work}

Our study has several limitations.
First, while we used LLM-based classifications with human validation to identify denial narratives, annotation accuracy is not perfect. Future research can develop more nuanced narrative identification methods.
Second, our findings are empirical and descriptive.
While we document temporal shifts of denial narratives and their influences, our analyses do not reveal the underlying psychological motivations that drive content creators to disseminate denialist information.

Third, the engagement metrics we analyze, such as video views, likes, and comment counts, provide indirect signals of audience interaction and do not directly reflect belief endorsement or attitude change among viewers.
Disentangling the high views of videos from algorithmic influences such as recommendation systems presents a natural future research step.
Finally, we observe that old denial narratives have diminished in influence over time, but our research does not examine the role of misinformation debunking efforts in driving this decline.
Future research can further investigate and quantify the effectiveness of debunking across different denial narratives.

\section{Conclusion}

In this work, we analyze the main themes and characteristics of climate denial videos on Brazilian YouTube, documenting a distinctive shift from ``old denial'' (scientific rejection) to ``new denial'' (solution-skeptic narratives).
Our findings indicate that new denial has not only outpaced old denial to become the dominant paradigm but has also transitioned from a personality-driven phenomenon to a decentralized network of actors.
New denial employs a sophisticated rhetorical approach designed to resonate with audiences who accept climate science but remain skeptical of climate action.
The evolution of these narratives raises important questions for the future of climate communication, particularly as generative AI threatens to lower the barriers for producing misleading content at scale. 
Current content moderation rules appear insufficient to address this shifting landscape, raising fundamental concerns about platform accountability. 
This governance gap illustrates how the evolving nature of climate misinformation can outpace platform policies, allowing harmful content to persist unchecked.

\section*{Generative AI Usage Statement}

Generative AI tools were not used for this manuscript in addition to checking typos. 
Claude Sonnet 4.5 was used solely for grammar checking parts of the text. GitHub CoPilot (1.388.0) was used to assist in coding in Visual Studio Code.
Generative AI (GPT-4.1-mini and Qwen3-30b) was used for text classification as described in the methodology section, being validated against human annotations to ensure performance.
The authors retain full responsibility for the originality and accuracy of all content in this work.

\section*{Ethical Considerations Statement}

In this work, we conduct a large scale analysis of climate denial on Brazilian YouTube as a representative country for the Global South. We recognize that doing so risks oversimplifying the issue that may manifest differently in different cultural contexts. For this reason, we explicitly acknowledge these limitations and avoid generalizing the findings beyond the current setting. 
Furthermore, this study consists of only publicly available YouTube data, consisting of video metadata and comments. All examples quoted are illustrative and are not intended to target specific individuals, and we do not share any more personal information than necessary. We also avoid labeling content as disinformation, since doing so assumes intent on the part of the video creator. Furthermore, all individuals cited in this research are known high-profile public figures in Brazil.
Finally, our findings are intended to support researchers, policy-making and platforms in understanding emerging climate misinformation strategies, with potential application for content moderation and public-interest interventions. We recognize that our findings may be exploited by malicious actors for the creation of new denialist content, but we believe that the benefits of understanding this already rapidly growing strain of denialism outweigh possible negative impacts.

\section*{Positionality Statement}
Five of the seven authors are based in Brazil and affiliated with a Brazilian public university (UFMG), including two professors, two M.Sc. students, and one B.Sc. student. These authors have direct experience with the Brazilian social media, political context, and climate discourse examined in this study. All annotations of Portuguese-language content were performed by native Brazilian Portuguese speakers within the team. We acknowledge that our academic background may shape how we interpret results.

\section*{Acknowledgments}

This project was supported by the Microsoft Accelerate Foundation Models Research (AFMR) program.
This research was supported by the National Research Foundation of Korea (RS-2022-00165347).
This work was partially funded by CNPq, CAPES, FAPEMIG, and IAIA-INCT on AI.
The authors thank anonymous reviewers for their insightful feedback and comments.


\bibliographystyle{ACM-Reference-Format}
\bibliography{0_main}


\appendix

\begin{figure*}[tb]
    \centering
    \includegraphics[width=\linewidth]{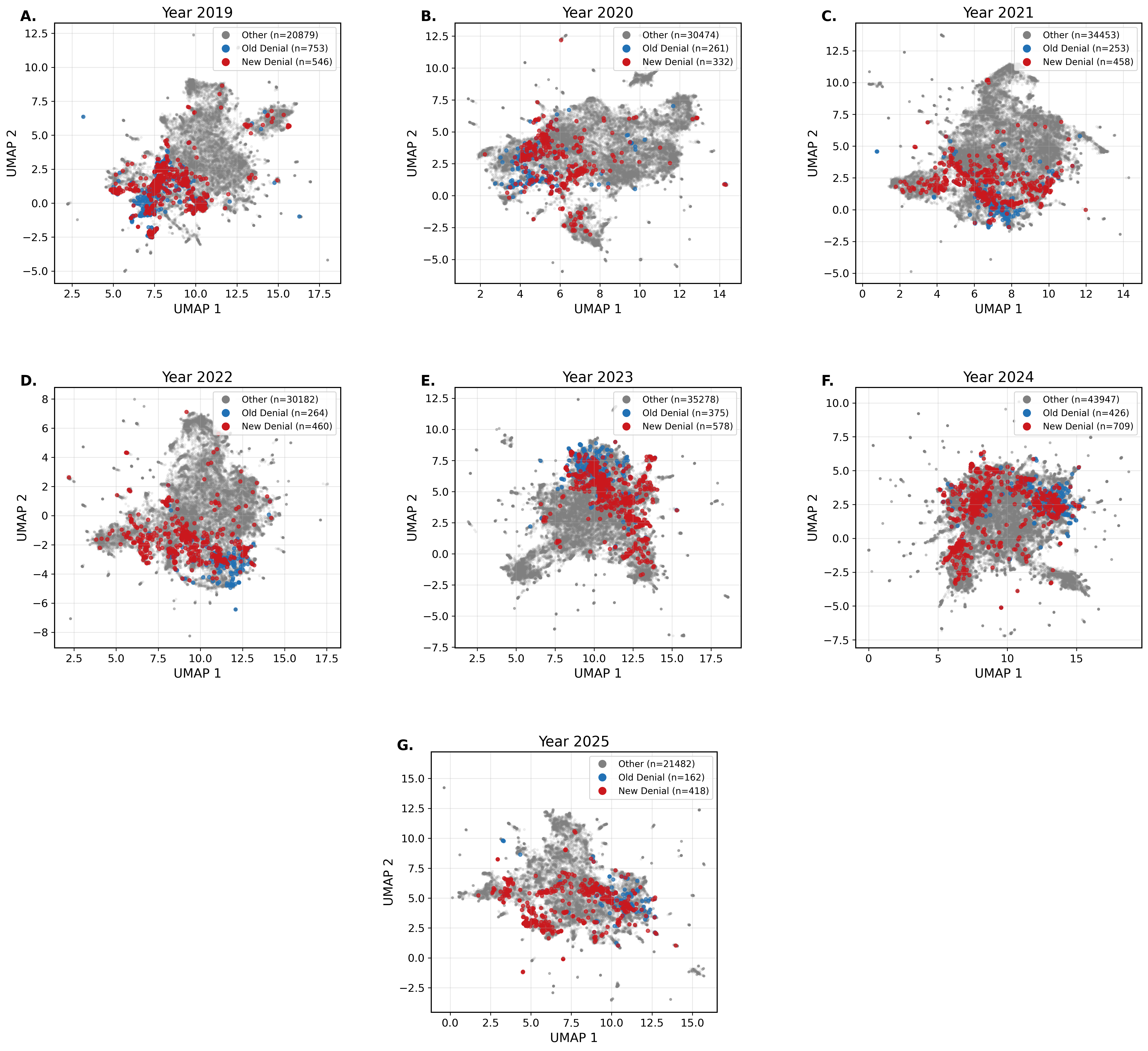}
    \caption{UMAP visualization of video content embeddings for each year, showing the semantic clustering of old denial (blue), new denial (red), and other climate-related content (gray). Note that in plot A, we changed the order of the new denial and old denial distributions from Fig.~\ref{fig:denial_analysis_stats} to show the high overlap between these two groups.}
    \label{fig:app_denial_all_embeddings}
    \Description[New denial became more different from old denial over time.]{Video embeddings for 2019 show that old and new denial were initially similar, but by 2024 new denial had changed remarkably.}
\end{figure*}

\begin{figure*}[tb]
    \centering
    \includegraphics[width=\linewidth]{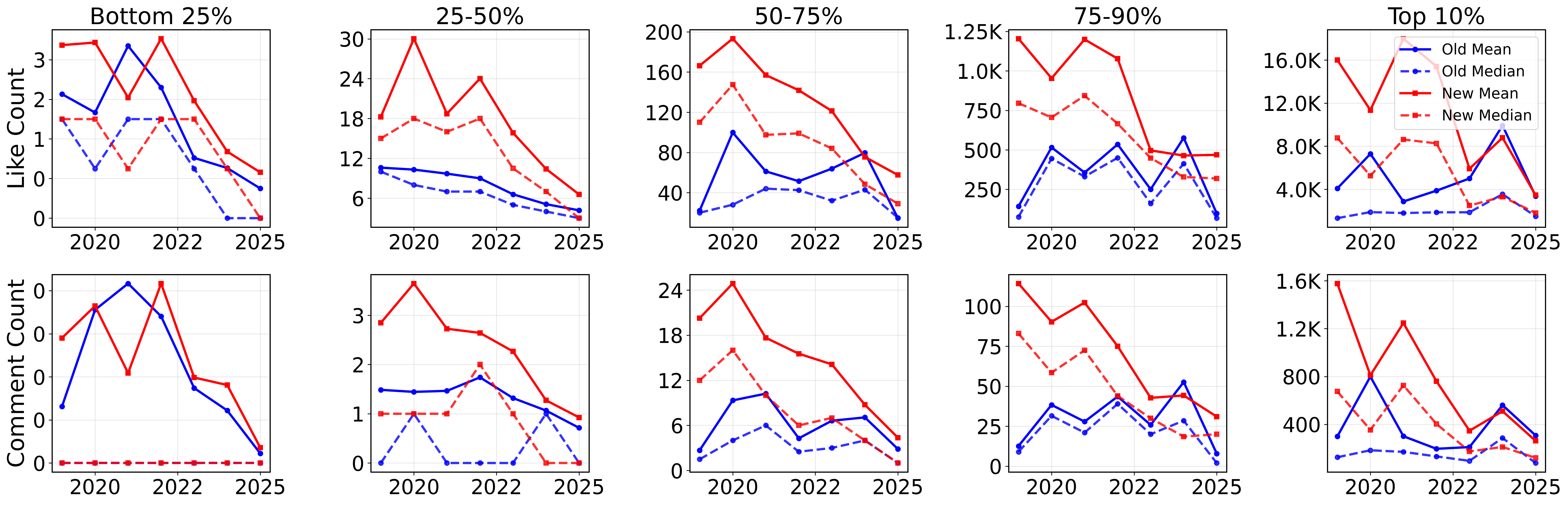}
    \caption{Yearly engagement as like and comment count for old denial (blue) and new denial (red) videos across quantile groups from 2019 to 2025. Solid lines represent mean values and dashed lines represent median values.}
    \label{fig:denial_engagement_like_comment}
    \Description[Engagement metrics for denial.]{Engagement metrics for denial. New denial metrics are generally higher than old denial.}
\end{figure*}

\section{Data Description}

Table \ref{tab:statistics} shows the statistics for each category in our dataset. In general, new denial video garner more impressions than their old denial counterparts. The vast majority of videos is not denialist. Notice how in general, new denial gets more impressions on average, in terms of likes, views and comments.

\begin{table*}[h]
\caption{Statistics for the dataset. Mdn. denotes the median. Cmt. denotes comments.}
\centering
\small
\label{tab:statistics}
\resizebox{\linewidth}{!}{%
\begin{tabular}{lrrrrrrrr} \toprule
                    & \multicolumn{1}{l}{\textbf{Videos}} & \textbf{Channels} & \textbf{Mdn.Views} & \textbf{Mean Views} & \textbf{Mdn.Likes} & \textbf{Mean Likes} & \textbf{Mdn.Cmt.} & \textbf{Mean Cmt.} \\ \midrule
\textbf{Old Denial} & 2,495                               & 1,390             & 137                 & 7,835               & 14                  & 727                 & 1                     & 53                     \\
\textbf{New Denial} & 3,503                               & 1,994             & 277                 & 12,495              & 23                  & 1,199               & 2                     & 86                     \\
\textbf{Others}     & 221,466                             & 82,471            & 146                 & 17,237              & 8                   & 484                 & 0                     & 17                    \\ \bottomrule
\end{tabular}
}
\end{table*}

Table \ref{tab:top_channels} shows the statistics for each channel category in our dataset. For both old and new denial, independent content creators make up the bulk of the content produced, closely followed by independent digital media. As expected, the more established groups such as government, research and education institutes are less associated with denialist rhetoric. In fact, many of the videos classified as denialist in these categories consist of recordings of speeches by specific politicians that could have denialist tone, not necessarily representing the views of the channel.

\begin{table*}[tb]
\centering
\caption{Statistics for channel categories in our dataset.}
\label{tab:top_channels}
\resizebox{\linewidth}{!}{%
\begin{tabular}{lrrrr} \toprule
                            & \multicolumn{2}{c}{\textbf{Old Denial}}                & \multicolumn{2}{c}{\textbf{New Denial}}                \\ \midrule
              \textbf{Channel Category}                       & \textbf{Video Count} & \textbf{ Total Subscribers} & \textbf{Video Count} & \textbf{Total Subscribers} \\ \midrule
Individual content creators          & 1,762                & 82,231,557                      & 1,971                & 124,457,008                     \\
Independent digital media            & 200                  & 40,724,257                      & 505                  & 79,666,044                      \\
Individual educators                 & 184                  & 8,892,247                       & 203                  & 12,495,376                      \\
Religious or spiritual organizations & 104                  & 4,677,126                       & 138                  & 5,738,234                       \\
Traditional news outlets             & 103                  & 82,527,096                      & 343                  & 104,715,400                     \\
Non-profit organizations             & 32                   & 389,652                         & 84                   & 766,774                         \\
Industry representatives             & 19                   & 59,599                          & 45                   & 404,972                         \\
Commercial companies                 & 18                   & 20,389,241                      & 62                   & 2,183,618                       \\
Formal education organizations       & 14                   & 93,869                          & 31                   & 883,082                         \\
National government                  & 10                   & 4,160,000                       & 31                   & 5,786,720                       \\
Research institutes                  & 6                    & 9,845                           & 21                   & 305,436                         \\
Political party                      & 3                    & 159,623                         & 28                   & 1,585,355                       \\
Local government                     & 1                    & 0                               & 1                    & 142,000                         \\
International organizations          & 0                    & 0                               & 1                    & 10,900                          \\
Other                                & 39                   & 772,083                         & 39                   & 39,291         \\ \bottomrule                
\end{tabular}
}
\end{table*}

\section{Topic Visualization}

\begin{figure*}[h]
     \centering
     
     \begin{subfigure}[b]{0.45\textwidth}
         \centering
         \includegraphics[width=0.9\textwidth]{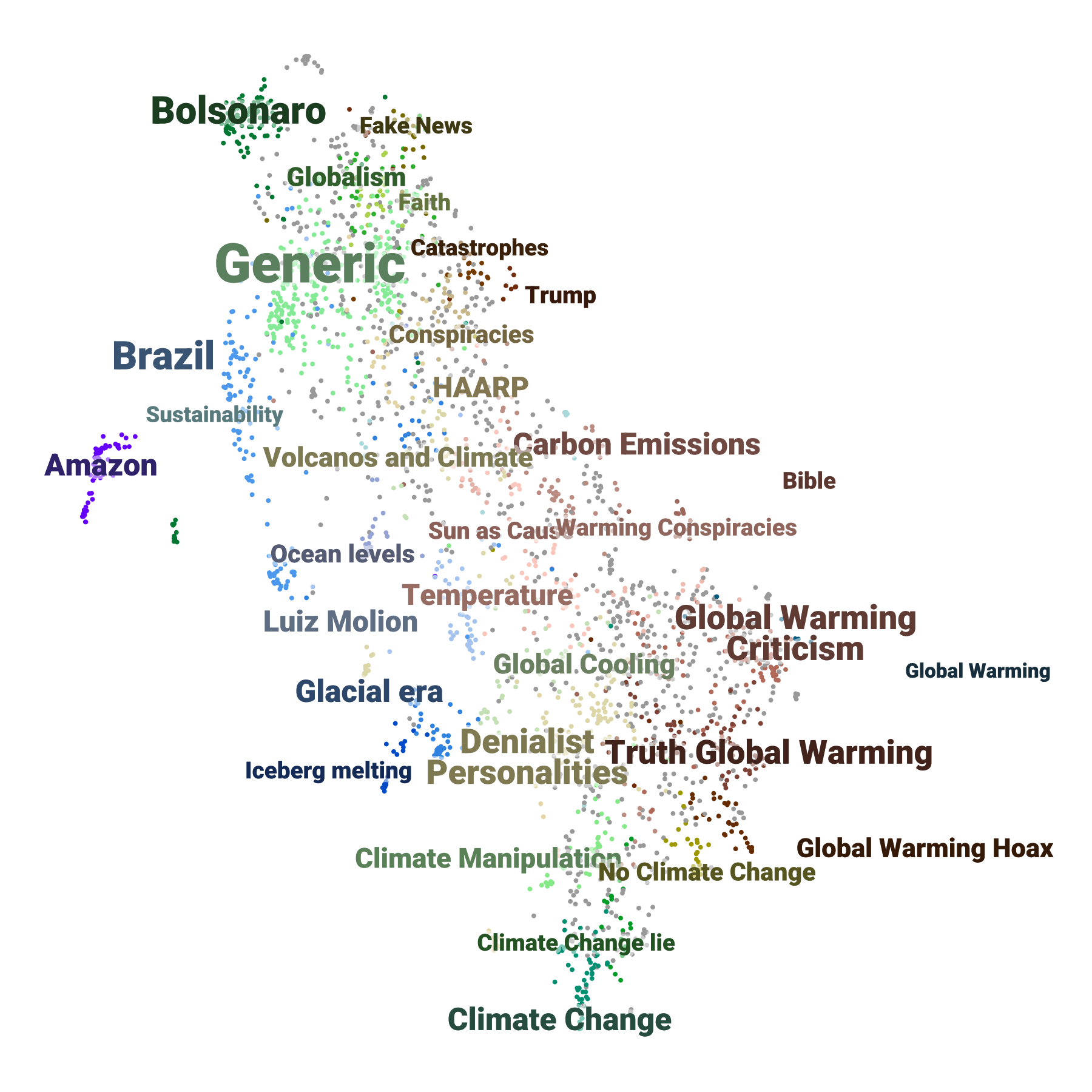}
     \end{subfigure}
     \hfill
     \begin{subfigure}[b]{0.45\textwidth}
         \centering
         \includegraphics[width=0.9\textwidth]{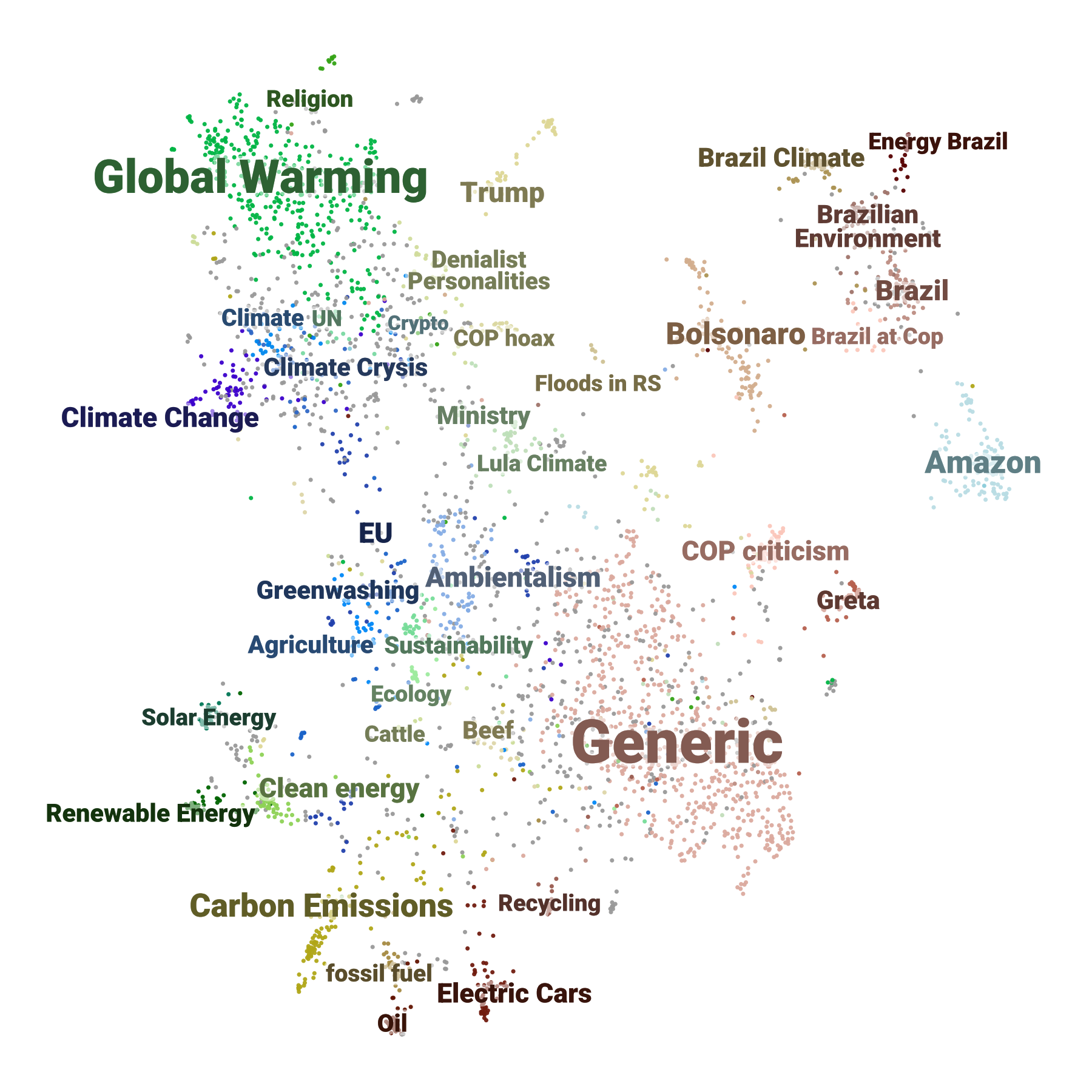}
     \end{subfigure}
     \caption{Two-dimensional visualization of topic modeling results using UMAP for dimensionality reduction.}
     \label{fig:topic_modeling_2d}
     \Description[Visualization of old and new denial topics.]{Visualization of old and new denial topics. Old denial is generally less well separated, possibly due to less topic variety, new denial displays some topics in close proximity, particularly those relating to climate in the top left corner, to Brazil in the top right and energy/economy in the bottom left.}
\end{figure*}

Figure~\ref{fig:topic_modeling_2d} shows the 2 dimensional visualization of the topics for old (left) and new denial (right) respectively. Note how some of the separations by major themes are visible in the new denial figure, as  climate related topics are concentrated at the top of the picture, while energy and economy topics are at the bottom left, with solutions occupying the center.

\section{Annotation Details}
\label{appendix:prompt}

We use two distinct prompts for this research. The first is used to create pseudo denial labels using Qwen3-30b. These labels were used simply to create a more balanced dataset for measuring the performance of our final classification. Table \ref{tab:prompt_qwen} contains the full user prompt.

\begin{table*}[ht]
    \caption{Prompt used to create pseudo stance labels using Qwen3-30b for a more balanced dataset.}
    \label{tab:prompt_qwen}
    \centering
    \small
    \begin{tabular}{|p{0.95\textwidth}|}\hline
You are a highly specialized system designed to analyze the characteristics of Brazilian Portuguese videos in the context of climate change and global warming discussions.

Your task is to analyze the provided Video Title, Video Description, and Video Transcript to determine two things:
1.  Is the video \textbf{relevant} to the topics of global warming or climate change?
2.  If it is relevant, what is the \textbf{stance} the video takes towards the issue?

\#\#\# \textbf{Stance Categorization Criteria}

You must categorize the video's stance into one of four categories following these precise criteria:

* \textbf{Belief:} The content acknowledges that \textbf{climate change and global warming are real} and are significantly influenced by \textbf{human activity} (anthropogenic causes).
* \textbf{Denial:} The content denies that human activity is the primary influence on climate change and global warming. This includes narratives such as:
    * "Global warming is not happening."
    * "Human-generated greenhouse gases are not causing global warming." / "Global warming is actually natural cycles or variation."
    * "The impacts of global warming are beneficial or harmless."
    * "Climate solutions won't work or are too costly."
    * "Climate science and the climate movement are unreliable/fabricated."
* \textbf{Conspiracy:} The content contains alternative, logically unsound, or pseudoscientific theories for what may be influencing climate change or global warming. Examples include:
    * "Inversion of magnetic fields", "Nibiru", "the beginning of a modern glacial era", "the New World Order", "the HAARP project", or other similar theories.
* \textbf{Neutral:} The content discusses climate change or global warming but \textbf{does not contain any explicit opinion} relating to human influence on these phenomena, neither affirming nor minimizing human contribution. This category is for purely informational or reportorial content without a clear opinionated stance.

\#\#\# \textbf{Input Data}

\textbf{Video Title:}{Title}
\textbf{Video Description:} {Description}
\textbf{Video Transcript:} {Transcript}

\#\#\# \textbf{Output Instruction}

Your response \textbf{must be only a single, valid JSON object} in the format specified below.

* If the video is not relevant to climate change or global warming, set `"relevant\_to\_climate\_change"` to `False` and set `"stance"` to `Neutral`.
* If the video is relevant, set `"relevant\_to\_climate\_change"` to `True` and set `"stance"` to the determined category (`Belief`, `Denial`, `Conspiracy` or `Neutral`).

```json
\{
    "relevant\_to\_climate\_change": Boolean, // True or False
    "stance": String // Belief, Denial, Conspiracy, or Neutral
\}
```\\ \hline
    \end{tabular}

\end{table*}

Table~\ref{tab:prompt_gpt} shows the system prompt for GPT-4.1-mini , which was used to generate the results reported in the paper, with the user prompt consisting of the video title, description and transcript. This prompt focuses on discerning between old and new denial, as that is one of the main interests of our analysis. However, as a side effect, the outputs from the model can also be used for identifying denialist content: videos that are marked as other and not as any of the narratives are not climate denial. GPT-4.1.-mini experiments were run with temperature set to 0 and top p = 1, and Qwen3-30B experiments were run with temperature set to $10^{-5}$ and top p = 0.95 to control for randomness. 

\begin{table*}[ht]
    \caption{Prompt used for classifying samples into old or new denial with GPT-4.1-mini.}
    \label{tab:prompt_gpt}
    \centering
    \small
    \begin{tabular}{|p{0.98\textwidth}|}\hline \\
    You are a highly specialized system designed to analyze Brazilian Portuguese YouTube videos discussing \textbf{climate change and global warming}.

Your task is to analyze the provided video details and classify the content into the following three categories.
If multiple narratives are present, choose all that apply. The other category should be selected only if neither denial categories apply.

\#\# Classification Categories

\#\#\# \textbf{Old Denial}
Content that \textbf{denies or disputes the scientific reality or causes of climate change}, focusing on whether it exists or whether humans are responsible.

This includes narratives such as:
- \textbf{Temperature records are manipulated}
  - “Os termômetros foram deslocados para áreas urbanas para falsificar o aquecimento.”
  - “A NASA / INPE ajusta dados históricos para exagerar o aquecimento.”

- \textbf{Short-term weather used to deny long-term trends}
  - “Nevou este ano, então o aquecimento global não pode ser real.”
  - “O último inverno foi mais frio do que nos anos 1970.”

- \textbf{Misuse of geological or historical time scales}
  - “A Terra era mais quente na Idade Média.”
  - “Os dinossauros viveram em climas mais quentes sem influência humana.”

- \textbf{Natural forcing explanations}
  - “Vulcões emitem mais CO2 do que os humanos.”
  - “Ciclos oceânicos como El Niño explicam todo o aquecimento.”
  - “Raios cósmicos ou a órbita da Terra explicam as mudanças climáticas.”

- \textbf{Rejection of the greenhouse effect}
  - “O CO2 é uma fração muito pequena da atmosfera para ter efeito.”
  - “O efeito estufa já está saturado, mais CO2 não muda nada.”

- \textbf{Sea-level rise minimization}
  - “Os marégrafos não mostram aceleração do nível do mar.”
  - “O nível do mar sempre subiu desde a última Era do Gelo.”

\#\#\# \textbf{New Denial}
Content that \textbf{accepts or bypasses the reality of climate change}, but argues \textbf{against action}, mitigation, or trust in climate science and institutions.

This includes narratives such as:
- \textbf{It’s not bad / impacts are exaggerated}
   - “A humanidade vai se adaptar sem problemas.”
   - “O frio mata mais pessoas do que o calor.”
   - “O aquecimento global vai abrir novas áreas agrícolas.”
   - “Os prejuízos de desastres aumentam só porque há mais pessoas e bens.”

- \textbf{Climate solutions won’t work}
   - “Energias renováveis dependem de combustíveis fósseis.”
   - “Painéis solares e turbinas eólicas não podem ser reciclados.”
   - “Carros elétricos apenas transferem a poluição para outro lugar.”
   - “As emissões do Brasil não fazem diferença no clima global.”
   - “China e Índia poluem mais, então não adianta agir.”

- \textbf{Climate policies are harmful}
   - “Leis climáticas servem para controlar os pobres.”
   - “Ambientalismo é contra o desenvolvimento.”
   - “Políticas climáticas vão destruir a agricultura e a indústria.”
   - “Imposto de carbono é apenas mais uma forma de roubo.”

- \textbf{Climate science, institutions or activists are unreliable}
   - “Cientistas do clima exageram para conseguir financiamento.”
   - “A mídia usa o medo para manipular a população.”
   - “Não existe consenso científico de verdade.”
   - “Mudança climática é ideologia política, não ciência.”
   - “A agenda climática é imposta por elites globais / ONU / Fórum Econômico Mundial.”

\#\#\# \textbf{Other}
Content that:
- Explicitly \textbf{acknowledges climate change as real and primarily human-caused}, \textbf{or}
- Does \textbf{not express a denialist or conspiratorial position}, including neutral, descriptive, journalistic, or unrelated content.

\#\# Analysis Process

Before assigning the final label, follow these steps:

1. \textbf{Reasoning}  
   Identify keywords, claims, or rhetorical strategies that indicate whether the video challenges:
   - the \textit{reality or causes} of climate change (\textbf{Old Denial}),
   - the \textit{need for action, solutions, or trust in science} (\textbf{New Denial}),
   - or neither (\textbf{Other}).

2. \textbf{Final Classification}  
   Select the most appropriate categories by marking \textbf{True} or \textbf{False} for each of the three categories:  
   \textbf{Old Denial}, \textbf{New Denial}, or \textbf{Other}.

\#\# Output Instruction

Your response \textbf{must be only a single, valid JSON object} in the following format:

```json 
\{ 
  "reasoning": String,
  "new\_denial": Boolean,
  "old\_denial": Boolean,
  "other": Boolean
\}
\\\hline
    \end{tabular}

\end{table*}

\end{document}